\documentclass[12pt]{iopart}

\usepackage[latin1]{inputenc}
\usepackage{graphicx,epsfig,color}
\usepackage[font=small]{caption}
\usepackage{wrapfig,rotating}

\begin{document}

\title{PHENIX measurements of 3D emission source functions in Au+Au collisions 
at $\sqrt{s_{NN}} = 200$ GeV }

\author{Roy A. Lacey for the PHENIX Collaboration}

\address{Department of Chemisry, Stony Brook University,
Stony Brook, NY 11794-3400, USA}

\ead{Roy.Lacey@Stonybrook.edu}

\begin{abstract}
A state-of-the-art 3D source imaging technique is used to extract 
the 3D two-pion source function in central and mid-central Au+Au 
collisions at $\sqrt{s_{NN}} = 200$~GeV. The source function indicates 
a previously unresolved non-Gaussian tail in the 
directions of the pion pair transverse momentum (out) and along the beam (long). 
Model comparisons give robust estimates for several characteristics of the emission 
source, including its transverse size, its mean proper breakup time $\tau$ and its 
emission duration $\Delta\tau$. 
These estimates are incompatible with the predictions for a first order phase 
transition. However, they point to significant relative emission times 
which could result from a crossover phase transition.

%
\end{abstract}


\section{Introduction}
		Lattice calculations indicate a rapid transition from a confined hadronic phase 
to a chirally symmetric de-confined quark gluon plasma (QGP) at the critical temperature 
$T_c \sim 170$~MeV \cite{Karsch:2000ps}. Such a plasma is produced in energetic Au+Au 
collisions at the Relativistic Heavy Ion Collider RHIC. However, it is currently 
unclear whether the phase transition to the QGP phase is first order ($\Delta T = 0$)
or a rapid crossover reflecting an increase in the entropy density 
associated with the change from hadronic ($d_H$) to quark and gluon ($d_Q$) 
degrees of freedom. 

	An emitting system which undergoes a first order phase transition 
is predicted to have a large space-time extent \cite{Pratt:1984su,Rischke:1996em}. 
This is because the transition ``softens'' the equation of state (ie. the sound speed $c_s \sim 0$) 
in the transition region, and this delays the expansion and considerably prolongs the 
lifetime of the system. A smaller space-time extent has been predicted for 
systems which undergo a crossover transition \cite{Rischke:1996em}.

\begin{wrapfigure}[20]{c}{0.5\textwidth}
\includegraphics[width=1.1\linewidth]{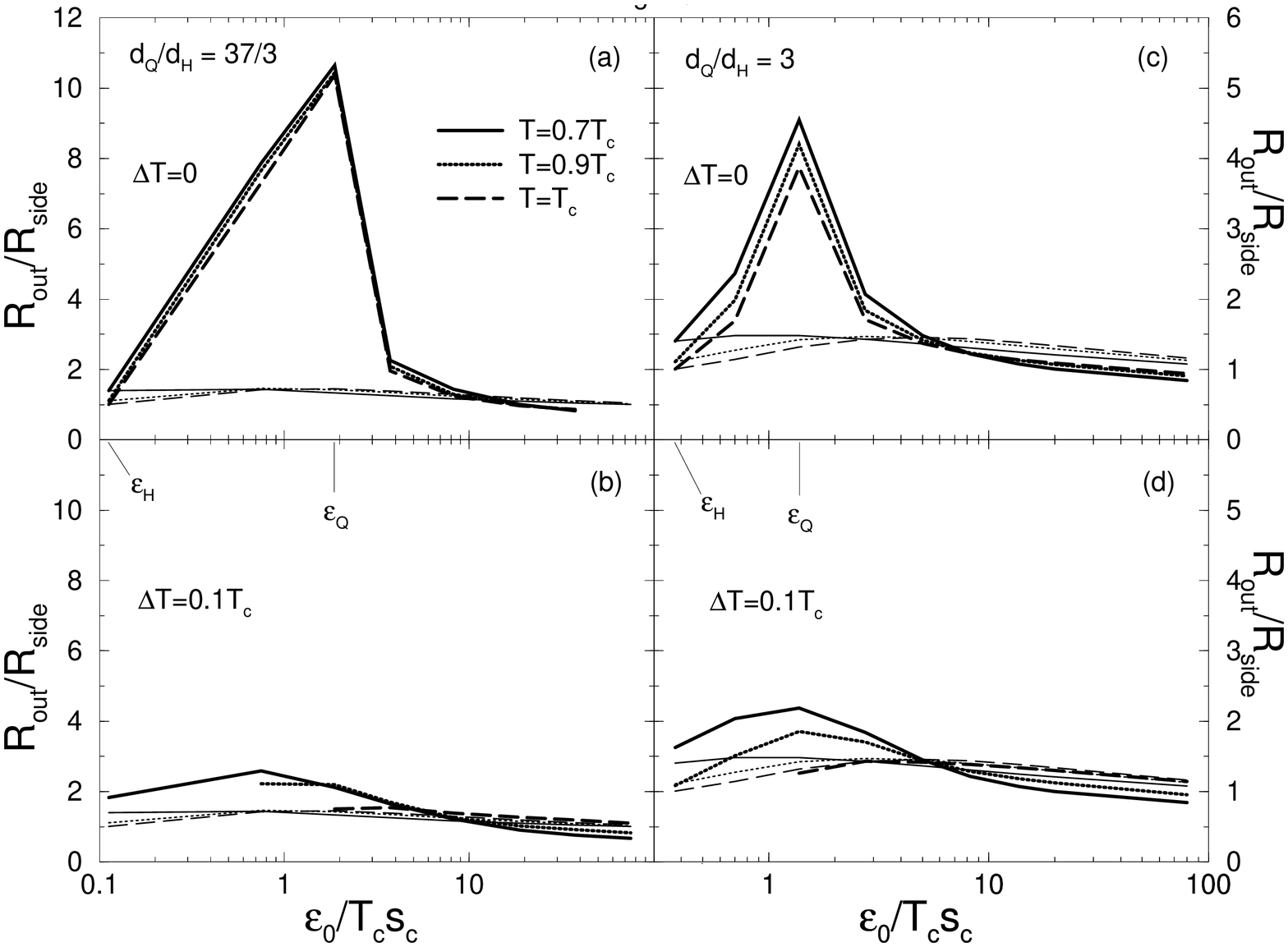} 
\caption{ $R_{out}/R_{side}$ as a function of the initial 
energy density for an expanding fireball. (a,b) are for $d_Q/d_H = 37/3$, 
(c,d) for $d_Q/d_H = 3$. The thick lines in (a,c) are for $\Delta T = 0$, in (b,d) for 
$\Delta T = 0.1T_c$. Thin lines show results for an ideal gas case. Solid lines show results 
for $T = 0.7T_c$, dotted for $T = 0.9T_c$, dashed for $T = T_c$. The figure is taken 
from Ref.~\cite{Rischke:1996em}.
\label{Fig1_Rischke_RoutRside}
}
\end{wrapfigure}

	To search for a prolonged lifetime, it has been a common 
practice to measure the widths ($R$) of the emission source 
function (assumed to be Gaussian) in the out- side- and 
long-direction ($R_{out}$, $R_{side}$ and $R_{long}$)
of the Bertsch-Pratt coordinate system \cite{Lisa:2005dd}. 
Here, the prediction is that $R_{out}/R_{side} >> 1$, for systems which 
undergo a first order phase transition, \cite{Pratt:1984su,Rischke:1996em}.
This is illustrated in Fig.~\ref{Fig1_Rischke_RoutRside}, 
where the values obtained from a hydrodynamical model 
calculation \cite{Rischke:1996em} are plotted as a function of 
energy density (in units of $T_cs_c$; $s$ is the entropy density).
%
%
These rather large ratios (cf. Figs.~\ref{Fig1_Rischke_RoutRside}a 
and c) have served as a major motivating factor for experimental 
searches at several accelerator facilities \cite{Lisa:2005dd}. 
No evidence for a prolonged lifetime were found by these studies and the reported 
Gaussian source functions are spheroidal with $R_{\rm out} \approx R_{\rm side}$ in 
the longitudinally co-moving system.

	A crossover transition can be modeled by varying the width $\Delta T$, of the 
transition region. Figs.~\ref{Fig1_Rischke_RoutRside}a and b show that the magnitude 
of $R_{out}/R_{side}$ is considerably reduced (by as much as a factor of four)
when calculations are performed for $\Delta T = 0.1T_c$. 
Thus, the space-time extent of the emitting system is significantly 
influenced by the cross over transition and 
this could lead to less prominent signals which require more sensitive methods of 
detection. Indeed, a recent study with a 1D source imaging technique has observed a long 
non-Gaussian tail in the radial source function and attributed it to possible lifetime 
effects~\cite{ppg52}.

	In this contribution, we report on recent efforts to study the 3D two-pion source 
function via a new state-of-the-art technique proposed by Danielewicz and 
Pratt~\cite{dan06}. Namely, the 3D correlation function is 
first decomposed into a basis of Cartesian surface-spherical harmonics to extract the 
coefficients, also called moments, of the expansion. They are then imaged or fitted with 
a trial function to extract the 3D source function, which is then used to probe the emission 
dynamics of the pion source \cite{chu08,ppg076}.

\section{Analysis Method}
The 3D correlation function $C(\mathbf{q}) = N_{\rm fgd}(\mathbf{q})/N_{\rm bkg}(\mathbf{q})$
was obtained by taking the ratio of the 3D relative momentum distribution for 
$\pi^+\pi^+$ and $\pi^-\pi^-$ pairs in the same event $N_{\rm fgd}(\mathbf{q})$ and  
those from mixed events $N_{\rm bkg}(\mathbf{q})$, where  
$\mathbf{q}=\frac{(\mathbf{p_1}-\mathbf{p_2})}{2}$ where $\mathbf{p_1}$ 
and $\mathbf{p_2}$ are the momentum 4-vectors in the pair center of mass system (PCMS). 

The 3D correlation function $C(\mathbf{q})$ was expanded in a 
Cartesian harmonic basis~\cite{dan06} to obtain the moments
\begin{equation}
C(\mathbf{q})-1 = R(\mathbf{q}) =\sum_{l, \alpha_1 \ldots \alpha_l}R^l_{\alpha_1 \ldots \alpha_l}(q) \,A^l_{\alpha_1 \ldots \alpha_l} (\Omega_\mathbf{q})
\label{eqn1}
\end{equation}
where $l=0,1,2,\ldots$, $\alpha_i=x, y \mbox{ or } z$, 
$A^l_{\alpha_1 \ldots \alpha_l}(\Omega_\mathbf{q})$
are Cartesian harmonic basis elements; ($\Omega_\mathbf{q}$ is the solid 
angle in $\mathbf{q}$ space); $R^l_{\alpha_1 \ldots \alpha_l}(q)$ are 
Cartesian correlation moments given by Eq.~(\ref{eqn2}); 
and $q$ is the modulus of $\mathbf q$.
\begin{equation}
 R^l_{\alpha_1 \ldots \alpha_l}(q) = \frac{(2l+1)!!}{l!}
 \int \frac{d \Omega_\mathbf{q}}{4\pi} A^l_{\alpha_1 \ldots \alpha_l} 
 (\Omega_\mathbf{q}) \, R(\mathbf{q}).
 \label{eqn2}
\end{equation}
\begin{wrapfigure}{r}{0.5\textwidth}
\includegraphics[width=1.0\linewidth]{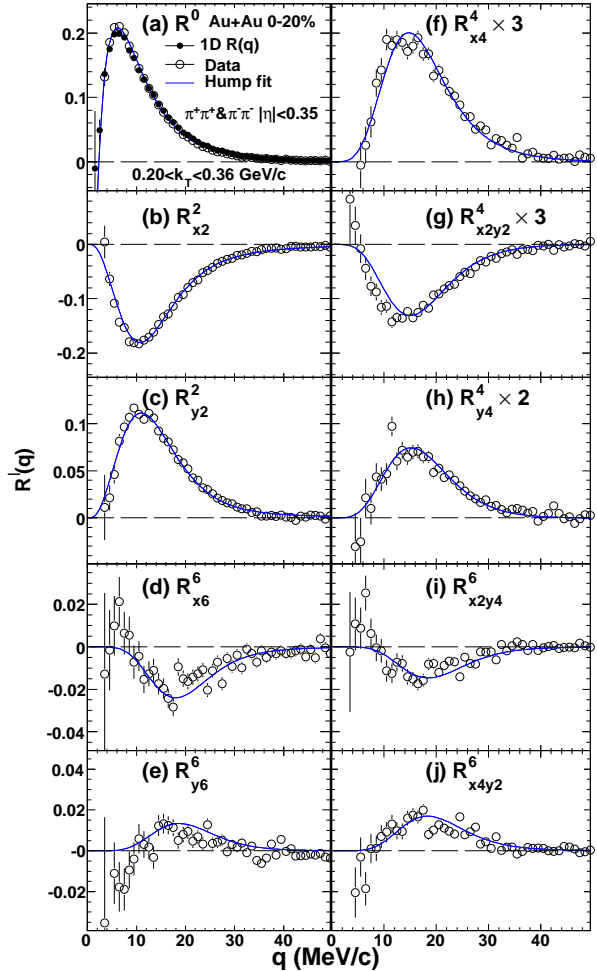}
 \caption{\label{phnx_fig1_ppg76}
{\small Experimental correlation moments $R^l(q)$ for $l=$0, 2, 4, 6. Panel (a) also 
shows a comparison between $R^0(q)$ and $R(q)$. Systematic errors are 
less than the statistical errors. The solid curves indicate
the Hump function Eq.~(\ref{hump_eqn}) fit.}}
\end{wrapfigure}
%

Here, the coordinate axes are oriented so that $z$ (long) is parallel to the beam 
direction and $x$ (out) points in the direction of the total transverse momentum of the pair.
For this analysis, Eq.~(\ref{eqn1}) was truncated at $l=6$ and expressed 
in terms of its 10 independent even moments: 
$R^0$, $R^2_{x2}$, $R^2_{y2}$, $R^4_{x4}$, $R^4_{y4}$, $R^4_{x2y2}$, 
$R^6_{x6}$, $R^6_{y6}$, $R^6_{x4y2}$ and $R^6_{x2y4}$ ($R^2_{x2}$ is 
shorthand for $R^2_{xx}$ \cite{dan06}); odd moments were checked and found to 
be consistent with zero [within statistical uncertainty] as required by symmetry 
considerations; 
higher order moments (for $l>6$) were also checked and found to be negligible. 

The independent moments are shown as a function of $q$ in Fig.~\ref{phnx_fig1_ppg76}. They 
were obtained by fitting the truncated series to the measured 3D correlation function with 
the moments as the parameters of the fit. 
It is noteworthy that the very good agreement between $R^0(q)$ and $R(q)$ (shown in panel (a)) 
points to the absence of any significant angular acceptance issues. It also attests 
to the reliability of the moment extraction technique. $R^0(q)$ and $R(q)$ 
both represent angle-averaged correlation functions, but $R^0(q)$ is obtained from 
the 3D correlation function via Eq.~(\ref{eqn2}) while $R(q)$ is evaluated directly 
from the 1D correlation function as in Ref.~\cite{ppg52}.

	The 3D source function $S(\mathbf{r})$ is obtained from the moments via imaging or fitting.
This is made transparent by the observation that, in analogy to Eq.~(\ref{eqn1}), $S(\mathbf{r})$ can 
also be expanded in a Cartesian Surface-spherical harmonic basis 
\begin{equation}
 S(\mathbf r) = \sum_l \sum_{\alpha_1 \ldots \alpha_l}
   S^l_{\alpha_1 \ldots \alpha_l}(r) \,A^l_{\alpha_1 \ldots \alpha_l} (\Omega_\mathbf{r}).
\label{eqn3}
\end{equation}
If the series for $R(\mathbf{q})$ and $S(\mathbf{r})$ are now substituted into the 
3D Koonin-Pratt equation;
\begin{equation}
C(\mathbf{q})-1 = R(\mathbf{q}) = \int d\mathbf{r} K(\mathbf{q},\mathbf{r}) S(\mathbf{r}),
\label{3dkpeqn}
\end{equation}
then, the following set of 1D 
expressions (Eq.~(\ref{momkpeqn}))~\cite{dan06} which relate the correlation 
moments $R^l_{\alpha_1 \ldots \alpha_l}(q)$ to the source moments $S^l_{\alpha_1 \ldots \alpha_l}(r)$
are obtained;
%
\begin{equation}
R^l_{\alpha_1 \ldots \alpha_l}(q) = 4\pi\int dr r^2 K_l(q,r) 
S^l_{\alpha_1 \ldots \alpha_l}(r).
\label{momkpeqn}
\end{equation}
%
Here, it should be noted that $S(\mathbf{r})$ gives the probability of emitting 
a pair of particles with a separation vector $\mathbf{r}$ in the PCMS and the 3D Kernel, 
$K(\mathbf{q},\mathbf{r})$, incorporates both the Coulomb force and the 
Bose-Einstein symmetrization. 

	The 1D imaging code of Brown and Danielewicz~\cite{bro97_98_01} was used to numerically 
invert each correlation moment $R^l_{\alpha_1 \ldots \alpha_l}(q)$ to extract the corresponding source 
moment $S^l_{\alpha_1 \ldots \alpha_l}(r)$. The latter were then combined as in Eq.~(\ref{eqn3}), to 
give the source function. The 3D source function was also extracted via direct fits to  
the 3D correlation function with an empirical Hump function given by
{\small
\begin{equation}
  \rm{S}^H(r_x,r_y,r_z) = \lambda \exp[-f_{\rm{s}}(\frac{x^2}{4 r_{x\rm{s}}^2} + 
  \frac{y^2}{4 r_{y\rm{s}}^2} + \frac{z^2}{4 r_{zs}^2}) 
  - f_{\rm{l}}(\frac{x^2}{4 r_{x\rm{l}}^2} + \frac{y^2}{4 r_{y\rm{l}}^2} + \frac{z^2}{4 r_{z\rm{l}}^2})], 
  \label{hump_eqn}
\end{equation}
}
where $\lambda, r_0, r_{\rm{xs}}, r_{\rm{ys}}, r_{\rm{zs}}, r_{\rm{xl}}, r_{\rm{yl}}, r_{\rm{zl}}$ 
are fit parameters and $f_{\rm{s}} = 1/[1 + (r/r_0)^2]$, $f_{\rm{l}} = 1 - f_{\rm{s}}$.
This procedure corresponds to a simultaneous fit of the ten independent moments. 
The solid curves in Fig.~\ref{phnx_fig1_ppg76} show the results of such a a fit. They 
indicate that the 8-parameter Hump function achieves a good fit to the data ($\chi^2/$ndf=1.4).
\section{Source Image and its interpretation}
\begin{wrapfigure}[20]{r}{0.5\textwidth}
\includegraphics[width=1.0\linewidth]{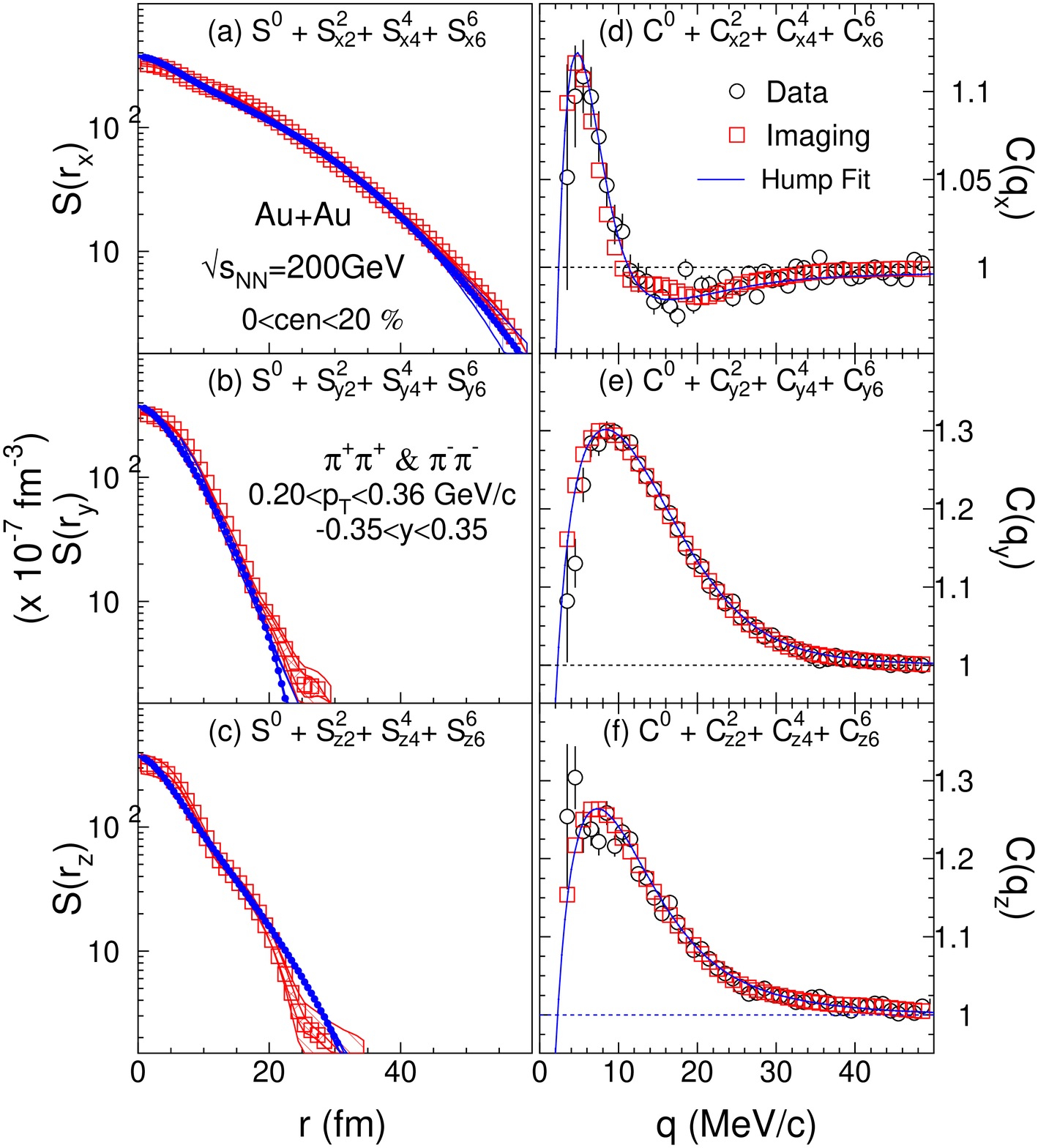}
\vskip -0.8cm  
 \caption{\label{phnx_fig2_ppg76}
{\small Source function profiles S$(r_x)$, S$(r_y)$ and S$(r_z)$ (left panels) and their 
associated correlation profiles C$(q_x)$, C$(q_y)$ and C$(q_z)$ (right panels) in the PCMS. 
The bands indicate statistical and systematic errors.}}
\end{wrapfigure} 
Figures~\ref{phnx_fig2_ppg76}(a)-(c) show the source function 
profiles S$(r_x)$, S$(r_y)$ and S$(r_z)$ obtained via fitting (line) and 
source imaging (squares). 
S$(r_x)$ is characterized by a long tail, which is resolved up 
to $\sim$60~fm; S$(r_y)$ and S$(r_z)$ are resolved up 
to $\sim$25~fm. The corresponding correlation profiles obtained by 
summation of the data (circle), fit (line) and 
image (square) moments are shown in Figs.~\ref{phnx_fig2_ppg76}(d)-(f).
The broader S$(r_x)$ is associated with the narrower C$(q_x)$ 
(Fig.~\ref{phnx_fig2_ppg76}(a) and (d)), as expected. 

	The extended tail lies along the pair total transverse momentum. 
Thus, the relative emission times between pions, as well as the source geometry, 
will contribute to S$(r_x)$. The source lifetime contributes to the range of 
S$(r_z)$ and S$(r_y)$ reflects its mean transverse 
geometric size. The difference between S$(r_x)$ and 
S$(r_y)$ is thus driven by the combination of the emission time 
difference, freeze-out dynamics and kinematic Lorentz boost.  

\begin{wrapfigure}[21]{c}{0.5\textwidth}
\includegraphics[width=1.05\linewidth]{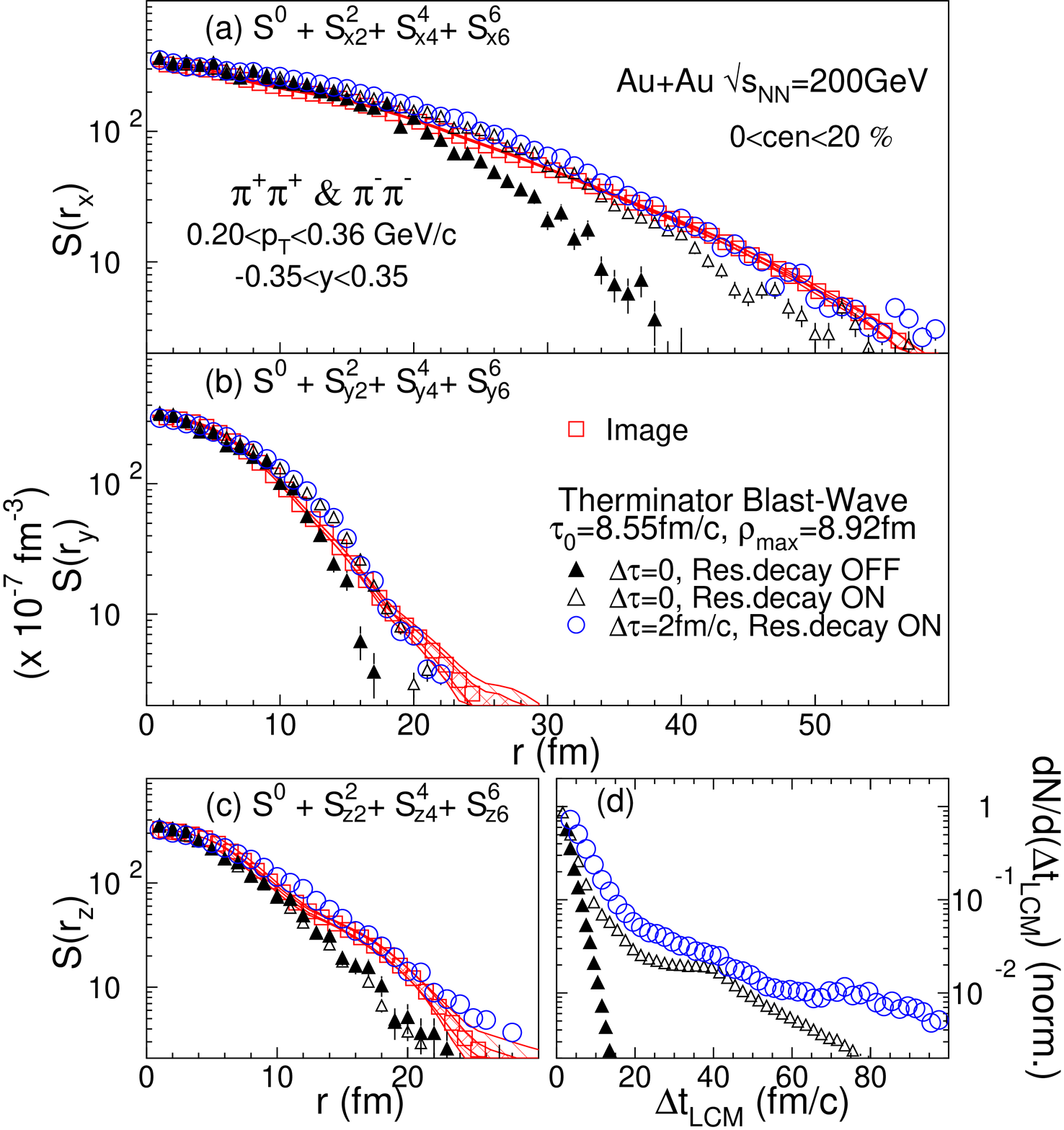}
\vskip -1.cm  
 \caption{\label{phnx_fig3_ppg76}
{\small Source function comparison between Therminator calculation and 
image for (a) S$(r_x)$, (b) $S(r_y)$, (c) $S(r_z)$ in PCMS. Panel (d) compares 
$\Delta \rm{t}_{\rm{LCM}}$ from Therminator events with various assumptions for 
$\Delta \tau$ and resonance emission.}}
\end{wrapfigure}
	To aid interpretation of the source function, the event generator 
Therminator~\cite{kis05} was used to predict source functions for several 
model scenarios. Therminator provides thermal emissions (including all 
known resonance decays) from a longitudinally oriented, boost 
invariant cylinder of radius $\rho_{\rm max}$. A differential fluid element 
is a ring defined by cylindrical coordinates $z$ and $\rho$; it breaks up at 
proper time $\tau$ in its rest frame or at time $t$ in the lab frame, 
where $t^2 = \tau^2 + z^2$. 
The freeze-out hypersurface is given by $\tau = \tau_0 + a\rho$, 
where $\tau_0$ is the proper breakup time for $\rho = $0 and $a$ represents 
the slope of the freeze-out hypersurface in $\rho$-$\tau$ space; for $a>0$ 
particles at small $\rho$'s are emitted at earlier times i.e inside-out ``burning''; 
outside-in ``burning'' occurs for $a<0$. 

	Calculations were performed in Blast-Wave 
mode for a set of parameters tuned to fit charged pion and kaon 
spectra. In addition, transverse expansion [governed by a radial 
velocity $v_r$ semi-linear in $\rho$, i.e. $v_r(\rho)=(\rho/\rho_{\rm max})/(\rho/\rho_{\rm max}+ 
v_t)$ ($v_t =1.41$)] was assumed and $a$ was varied. 
The mid-rapidity pion pairs so obtained were also generated with the effects 
of all known resonance decay processes on and off. 
The resulting pairs were then transformed to the PCMS, as in the data analysis, to 
obtain S$(r_{x,y,z})$ distributions for comparison with the data. 

Figure~\ref{phnx_fig3_ppg76} shows that the 3D source function generated by 
Therminator calculations (open triangles) with $\tau_0=8.55$~fm/c, $\rho_{\rm max}=8.92$~fm 
and $a=-0.5$ with resonances on, reproduce S($r_y$) but do not fully account for the long tails 
in $x$ and $z$ ie. the latter are longer than the Therminator source 
profiles. For the same parameters, the calculations underestimates S$(r_x)$, S($r_y$) and 
S$(r_z)$ when resonances are turned off (solid triangles). Reasonable attempts to fit the 
distributions by only increasing $\tau_0$ or with $a \ge 0$ failed, suggesting substantial 
contribution from pion pairs with significantly longer emission time differences. 
 
	An alternative approach to lengthen the distribution of time differences between pion 
pairs is to sample them from a family of hypersurfaces defined by a range of 
values of proper breakup times $\tau'$. One such parametrization consists 
of replacing $\tau$ by $\tau'$ chosen from an exponential distribution 
$dN/d\tau' = \frac{\Theta(\tau'-\tau)}{\Delta\tau} \exp[-(\tau'-\tau)/\Delta\tau]$, where the width 
of the distribution $\Delta\tau$ represents the mean proper emission duration.  
Figure~\ref{phnx_fig3_ppg76} shows  that this approach, 
with $\Delta\tau=2$~fm/c (open circles), leads to a fairly good match 
to the three observed source profiles. 

Figure~\ref{phnx_fig3_ppg76}(d) summarizes the relative emission time distribution in 
the LCMS, $\Delta \rm{t}_{\rm{LCM}}$, for pion pairs from events with the parameterizations 
indicated. For a fixed $\tau_0=8.55$~fm/c ($\Delta \tau =0$) and 
resonance decays excluded, the distribution $\Delta \rm{t}_{\rm{LCM}}$ is narrow, 
$\left\langle |\Delta \rm{t}_{\rm{LCM}}| \right\rangle =2.4$~fm/c. The addition of 
resonance decays adds a long tail and gives 
$\left\langle |\Delta \rm{t}_{\rm{LCM}}|\right\rangle =8.8$~fm/c. Replacing $\tau$ with the 
exponential distribution $\tau'$ with $\Delta \tau=2$~fm/c, results in a 
$\Delta \rm{t}_{\rm{LCM}}$ distribution which is significantly broadened to give 
$\left\langle |\Delta \rm{t}_{\rm{LCM}}| \right\rangle =11.8$~fm/c. The wider distribution 
of time delays is needed to reproduce the source distributions. This implies a  
non-zero proper emission duration in the emission rest frame. 
 
	Figure~\ref{phnx_fig3_ppg76} shows that substantial time 
differences $\Delta \rm{t}_{\rm{LCM}}$ are required to account for the source distensions; 
however, the interplay between proper time and breakup dynamics is model dependent. 
Nevertheless, the picture which emerges from the Therminator model comparison 
is consistent with an expanding fireball ($\rho_{\rm max}=8.92$~fm) with proper breakup 
time $\tau_0 \sim 9$~fm/c, which hadronizes and emits particles over a short but 
non-zero mean proper emission duration $\Delta\tau = 2$~fm/c. Such a short 
time duration is incompatible with the predictions  \cite{Pratt:1984su,Rischke:1996em}
for a first order phase transition.

\section{Conclusions}
In conclusion, a novel three-dimensional source imaging 
technique has been used to extract the 3D pion emission source 
function in the PCMS frame from Au+Au collisions at $\sqrt{s_{NN}}=200$~GeV. 
The source function has a much greater extent in the 
out ($x$) and long ($z$), than in the side ($y$) direction. Therminator 
model comparisons suggest an emission source ($\rho_{\rm max}=8.92$~fm) burning from 
outside in with proper lifetime $\tau_0 \sim 9$~fm/c and a mean proper emission duration 
$\Delta\tau \sim 2$~fm/c. These emission characteristics are incompatible with the 
predictions for a first order phase transition. 
However, they point to significant relative emission times 
($\left\langle |\Delta \rm{t}_{\rm{LCM}}| \right\rangle\approx 12$~fm/c, 
including those due to resonance decay) which could result from a crossover 
phase transition.

\end{document}